\documentstyle[aps,multicol,pre,eqsecnum,epsfig]{revtex}

\newcommand{\equ}[1]{(\protect\ref{#1})}
\newcommand{\Generic}[3]{{#1}{ (#2)}{ #3}.}
\newcommand{\ep}{\varepsilon}
\newcommand{\al}{\alpha}
\newcommand{\cP}{{\cal P}}
\newcommand{\cI}{{\cal I}}
\newcommand{\Raya}{\noindent\makebox[3.4truein]{\hrulefill}}
\newcommand{\MycapS}[1]{\protect\noindent\protect\parbox{8.6cm}{\small
    #1}}

\begin{document}

\draft

\title{The Maximum Entropy principle and the nature of fractals}

\author{R. Pastor-Satorras\footnote{Present address: Dept.  of Earth,
    Atmospheric, and Planetary Sciences, MIT, Cambridge, MA
    02139-4307.\\ E-mail:{\tt romu@segovia.mit.edu}}}

\address{Departament de F\'{\i}sica Fonamental, Facultat de
F\'{\i}sica, Universitat de Barcelona, Diagonal 647, E-08028 Barcelona
(Spain)}

\author{J. Wagensberg}

\address{Centre d'Estudis de la Complexitat, Museu de la Ci\`{e}ncia de
la Fundaci\'{o} ``la Caixa'', Teodor Roviralta 55, E-08022 Barcelona
(Spain)}

\maketitle

\begin{abstract}
  We apply the Principle of Maximum Entropy to the study of a general
  class of deterministic fractal sets. The scaling laws peculiar to
  these objects are accounted for by means of a constraint concerning
  the average content of information in those patterns.  This
  constraint allows for a new statistical characterization of fractal
  objects and fractal dimension.

\end{abstract}


\begin{multicols}{2}

\section{Introduction}

The frequent emergence of fractal objects \cite{mandelbrot82,feder88}
in Nature is a very relevant issue in contemporary physics. Since the
publication of the celebrated work of B. B. Mandelbrot
\cite{mandelbrot82}, fractal geometry has ideed found a succesful use
in describing these patterns in physics~\cite{feder90} and natural
morphogenesis~\cite{vicsek94,dawkins96}. However, we still do
not know why fractals are so strikingly frequent and stable in Nature.

In a recent paper \cite{pastor96} a tentative argument was proposed in
order to give account of  the fractal nature of diffusion-limited
aggregation (DLA) clusters \cite{witten81}. In that paper, the
branching structure of DLA clusters was analyzed within a new
framework, in which the stress was laid not on the {\em order} of the
branches (as it was traditionally done \cite{hinrichsen89}) but on
their {\em mass}. In this approach, a new magnitud was introduced, the
{\em branch distribution} $n(s,M)$, defined as the average
number of branches of a given mass $s$ present in a typical cluster of
$M$ particles. From numerical simulations it was found that, in the
limit of large branches, this distribution is a scaling function of
$s$, namely $n(s, M)/M \sim s^{-\al}$.  This particular functional form
(power
law), which on the other hand turns out to be universal in fractal
sets, was accounted for by the {\em Maximum Entropy} (MaxEnt)
principle.

The Theory of Information \cite{shannon48,brillouin62} and the Maximum
Entropy
formalism \cite{jaynes57}  are well-known methods in statistical
physics and time-honored tools for the study of complex systems,
especially in those stationary situations where one faces a
considerable lack of information. These techniques have found a wide
number of
applications, among which we can mention the prediction
of average magnitudes in statistical mechanics (where there is an
extreme lack of information) \cite{jaynes57}, the analysis of
signals hidden in noise \cite{marple87}, the prediction of ecological
stationary states \cite{wagens90}, models of growth and differentiation
\cite{wagens92}, etc. (For a general review of applications of the
MaxEnt Principle, see for instance Refs.~\cite{wagens90}
and~\cite{wagens92}.) 

Within the general approach of the MaxEnt formalism
\cite{pastor96,wagens92}, a complex system $X$ is represented by a set
of $n$ {\em nodes}, characterized by a certain
extensive magnitude $x$. The description of the system is
accomplished by means of a distribution of
{\em structural probabilities} which assigns a probability $p(x_i)$ to
each of the nodes $i=1,\ldots,n$.  In some cases, the structural
probabilities have a truly physical meaning; an example is provided in
the application of MaxEnt to ecological energy-flow networks
\cite{wagens90}, where the nodes represent compartments in a
biologically meaningful partition of the ecosystem and the structural
probabilities  represent fractions of energy exchange between different
nodes. In other cases, however, the structural probabilities have a
merely statistical interpretation. For instance,
$p(x_i)$ can represent the probability of node $i$ possesing a given
amount $x_i$ of the characteristic magnitude $x$; we clearly
recognize here the analogy with statistical mechanics, where the nodes
are the different states accessible to the system and the role of $x$
is played by the energy in those levels.
Whatever would be the interpretation of the
probabilities $p(x_i)$, the system $X$ is globally  characterized
by its total {\em entropy}, measured by the Shannon formula (expressed
in {\em nits}) \cite{shannon48,brillouin62}
\begin{equation}%
H(X) = - \sum_i p(x_i) \ln p(x_i).
\label{entropy}
\end{equation}%
On these bases, the MaxEnt principle as stated by Jaynes says that
\cite{pastor96,jaynes57,wagens92}: ``{\em The least biased and most likely
probability assignment $p(x_i)$ is that which maximizes the total
entropy $H(X)$ subject to the constraints imposed on the system}''. The
interpretation of this procedure is quite clear \cite{jaynes85}. The
magnitude~\equ{entropy} is a measure of the amount of
uncertainty \cite{brillouin62} in our description of the system $X$
through the
distribution $p(x_i)$. Therefore, the distribution that describes $X$
in a least biased way, given the information available about the
system and without assuming anything else, is that one which maximizes
the entropy $H(X)$, subject
to the constraints imposed by that very information available. Besides
the trivial normalization condition $\sum_{i=1}^n p(x_i) = 1$, these
constraints imposed on the system have to be understood as the global
effect of the fundamental laws involved in the process.

In this view, the branching structure of DLA discussed in
Ref.~\cite{pastor96} was resolved by imposing a constraint
over the average {\em generating information} (to be defined
later on) in the set of branches in the ensemble of all DLA clusters
of a given mass $M$. This generating information was interpreted as
the
information required to fully specify the structure of a
generic cluster, and is therefore related to the {\em algorithmic
complexity} \cite{chaitin75} of the DLA process.

The purpose of the present paper is to  propose a new generalized 
Maximum Entropy argument, in order to give reason of a wide class
of deterministic fractal structures that can be defined by an
{\em iterative process}. Examples of such kind of structures are
found in the construction  of fractal sets by means of {\em
Iteration Function Systems} \cite{barnsley88}. In section~2  we study a
simple model of a one-level iterative construction. We apply to it the
MaxEnt principle, using as a constraint a suitable expression for the
generating information, which yields as expected a power law
behaviour. An example of this model is analyzed, showing a
thermodynamic analogy which allows a new statistical interpretation of
fractal sets and fractal dimension. Section~3 generalizes the previous
situation, considering a more complex two-level model closely related
to
the DLA process discussed in Ref~\cite{pastor96}; an example is also
provided and analyzed.  Finally, we comment our conclusions in
section~4.

\section{One-level iterative model}

Let us first consider the  model of a simple iterative process $\cP$,
whose iteration through a given number of levels (labelled with index
$k$) leads to some final pattern.  In the $k$-th level we have a
structure composed of $n_k$ {\em identical} elements of order $k$;
$n_k$ is the {\em occupation number} of level $k$.  Each
one of those order $k$ elements is characterized exclusively by a
certain magnitude $\ell_k$, which we measure in units of a certain
{\em atom} $\ep$ (the resolution). If $\ell_k$ is a length,
then $\ep$ would correspond to the so-called lower cutoff length.
However, in order to allow for the possibility of considering any other
extensive magnitudes to characterize the levels, we will mantain
instead the term atom.

In order to apply the MaxEnt formalism, we need to associate to $\cP$
some meaningful structural probabilities. We choose a distribution that
assigns to each level $k$ a probability $p_k$ proportional to its
occupation number, $p_k=n_k / \sum_{k'} n_{k'}$. Given
the final structure, constructed through an increasing sequence of
nested levels, $p_k$ is  the weighted
probability of any element in the sequence belonging to order $k$. This
distribution is trivially normalized to $1$. The total entropy of
$\cP$ is then given by
\begin{equation}%
H(\cP)= - \sum_k p_k \ln p_k.
\label{entro1}
\end{equation}%
This function is a measure of the diversity of the iteration process
throughout its whole history, taking into account the population
density of each one of its levels.

We next identify the constraints imposed on the system. Let us consider
the very nature of the iterative process, which we assume that leads to
a
final fractal pattern. This fractal limit is essentially characterized
by its {\em self-similarity} \cite{mandelbrot82}.  This concept is
linked to the fact that the structure of the whole object is {\em
similar} (at least in some statistical sense) to that of a small part
of it.  In other words, since the whole can be reproduced by any small
part of it, then the amount of information ({\em generating
information}) needed to reconstruct the object (measured in some suitable
way) is small and nearly equal to the information contained in any
small
portion of it.  The situation is quite the same for the {\em Iteration
Function Systems} \cite{barnsley88}.  In that case, all the information
required for the construction a complex deterministic
fractal is compressed into a small number of contraction functions
and, in the simplest case, into a small set of real numbers.
Hence, these fractals do indeed contain a small amount of
information.

In order to express this information content, let us consider an
element of order $k$, characterized  by the magnitude
$\ell_k$.  This element is therefore composed of $N_k=\ell_k/\ep$ atoms
of size $\ep$, arranged in a certain way.  If atoms are considered
to be {\em indistinguishable}, then the amount of information needed to
specify the arrangement of those $N_k$ atoms in the element is the same
as that involved in the problem of selecting $N_k$ objects with
equal probability. Elementary information theory \cite{shannon48} tells
us that this information is $\ln N_k$.  Accordingly, the amount  of
generating information used up for the specification of an element of
order $k$ is $I_k = \ln N_k = \ln \left( \ell_k / \ep \right)$.  The
average information over the whole iteration process $\cP$ is then
\begin{equation}%
\left<I\right> = \sum_k p_k I_k = \sum_k p_k \ln \left(
\frac{\ell_k}{\ep} \right).
\label{lligam1}
\end{equation}%
At this point we should stress the fundamental difference between
magnitudes~\equ{entro1} and~\equ{lligam1}. Both are informations, in
the strict sense of the mathematical Theory of Information. However,
the
entropy~\equ{entro1} refers to the {\em diversity} of an iterative
process, while~\equ{lligam1} is a tentative measure of what
we have to know in order to reproduce that very iterative process.

Our fundamental assumption concerning $\cP$ is that the relation
$\left<I\right> = \bar{I}=\mbox{\rm const.}$ works as a constraint
acting on the process and that there are no further relevant
constraints playing any essential role. The MaxEnt principle is then
applied by maximizing the total entropy~\equ{entro1}, subject to the
constraint of average generating information~\equ{lligam1} constant.
The maximization is performed by means of the Lagrange multipliers
method \cite{landsberg78}, computing the extremes of the auxiliary
function

\end{multicols}

\widetext

\Raya\hfill

\begin{displaymath}%
\Phi= - \sum_k p_k \ln p_k +
\beta \left( \bar{I} - \sum_k p_k \ln \left( \frac{\ell_k}{\ep}
\right) \right) + \beta' \left( 1 - \sum_k p_k \right),
\end{displaymath}%

\hfill\Raya

\begin{multicols}{2}

\noindent where $\beta$ and $\beta'$ are Lagrange multipliers. The extreme
(more precisely, the maximum), given by the condition $\partial \Phi /
\partial p_k = 0$, yields to the equation
\begin{displaymath}%
-\ln p_k - 1  - \beta \ln \left( \frac{\ell_k}{\ep} \right)
-\beta' = 0,
\end{displaymath}%
from which we obtain the following expression for the structural
probabilities
\begin{displaymath}%
p_k = e^{-1-\beta'} \left( \frac{\ell_k}{\ep} \right)^{-\beta}.
\end{displaymath}%
This equation, together with the relation $p_k \sim n_k$, provides the
functional form for the occupation numbers
\begin{equation}%
n_k = \mbox{\rm const.} \times \left( \frac{\ell_k}{\ep}
\right)^{-\beta}.
\label{theoretical1}
\end{equation}%
In other words, the occupation numbers scale
as a power of $\ell_k$, which implies a {\em self-similar} behaviour of
those magnitudes with respect to that variable~\cite{mandelbrot82}. The
interpretation of $\beta$ is linked to the fractal dimension of the
final pattern, as will become clear later on.

Eq.~\equ{lligam1} can be readily interpreted in a statistical framework
by imposing the normalization condition $\sum_k p_k = 1$, yielding to
the expression
\begin{equation}%
p_k = \frac{\left( \ell_k / \varepsilon  \right)^{-\beta}}{\sum_{k'}
\left( \ell_{k'} / \varepsilon  \right)^{-\beta}}.
\end{equation}%
By introducing the {\em partition function}
\begin{equation}%
Z(\beta) =  \sum_k \left( \frac{\ell_k}{\varepsilon} \right)^{-\beta} =
\sum_k \exp \left( - \beta \ln \bigl( \ell_k / \varepsilon  \bigr)
\right),
\label{partition}
\end{equation}%
the structural probabilities are given by
\begin{equation}%
p_k = \frac{\left( \ell_k / \varepsilon  \right)^{-\beta}}{Z(\beta)}.
\end{equation}%
and the average information~\equ{lligam1} is given by
\begin{equation}%
\left< I \right> = - \frac{\partial}{\partial \beta} \ln Z(\beta).
\label{infor}
\end{equation}%
That is, the iterative process is analogous to a canonical ensemble
from statistical mechanics \cite{landsberg78},
defined by a ``spectrum'' of information levels ${\cal I}_k = \ln
\bigl( \ell_k / \varepsilon  \bigr)$. The partition function of this
ensemble is $Z(\beta)$, with a ``temperature'' $1/\beta$ which can be
computed in principle by eq.~(\protect\ref{infor}), if the actual
value $\bar{I}$ of the constraint is known.

As an example of this kind of structure, let us consider the
iterative construction of the van Koch curve
\cite{mandelbrot82,wagens92}, the first steps of which are depicted in
figure~1. The process uses two simple forms: An initial straight
segment of length $1$ (stage $k=0$), the {\em initiator}, and a
{\em generator}.  This latter is a broken line made up of $N$ segments
of equal length $r<1$.  Every iteration starts with a broken line, and
proceeds by replacing every rectilinear segment with a reduced copy of
the generator, rotated in such a way that it has the same extremes as
the interval being replaced. In the $k$-th level of the process, we
have a broken line made up of  $n_k=N^k$ equal straight segments of
length $\ell_k = r^k$, which we identify as the elements of that 
level $k$.  Iterating this procedure infinitely many times leads to a
strictely self-similar fractal curve with fractal dimension $D= \ln N /
\ln (1/r)$ \cite{mandelbrot82}.  On the other hand, if the iteration is
stopped after $m+1$ steps, we obtain a final structure composed of
elements of minimum length $\ell_m=r^m$, which we take as the atom (in
this case, the cutoff length) $\ep$, so that $\ell_k / \ep = r^k /
r^m$.  By taking logarithms in both expressions $n_k$ and $\ell_k /
\ep$ and removing the dependence on $k$ we get

\end{multicols}

\widetext

\Raya\hfill

\begin{equation}%
n_k = N^k = \left( r^k \right)^{-D} =
\left[ r^m \left( \frac{\ell_k}{\ep} \right)  \right]^{-D} =
(r^m)^{-D}  \left( \frac{\ell_k}{\ep} \right)^{-D} =
N^m \left( \frac{\ell_k}{\ep} \right)^{-D}.
\label{n-k-koch}
\end{equation}%

\hfill\Raya

\begin{multicols}{2}

\begin{figure}[t]
  \centerline{\epsfig{file=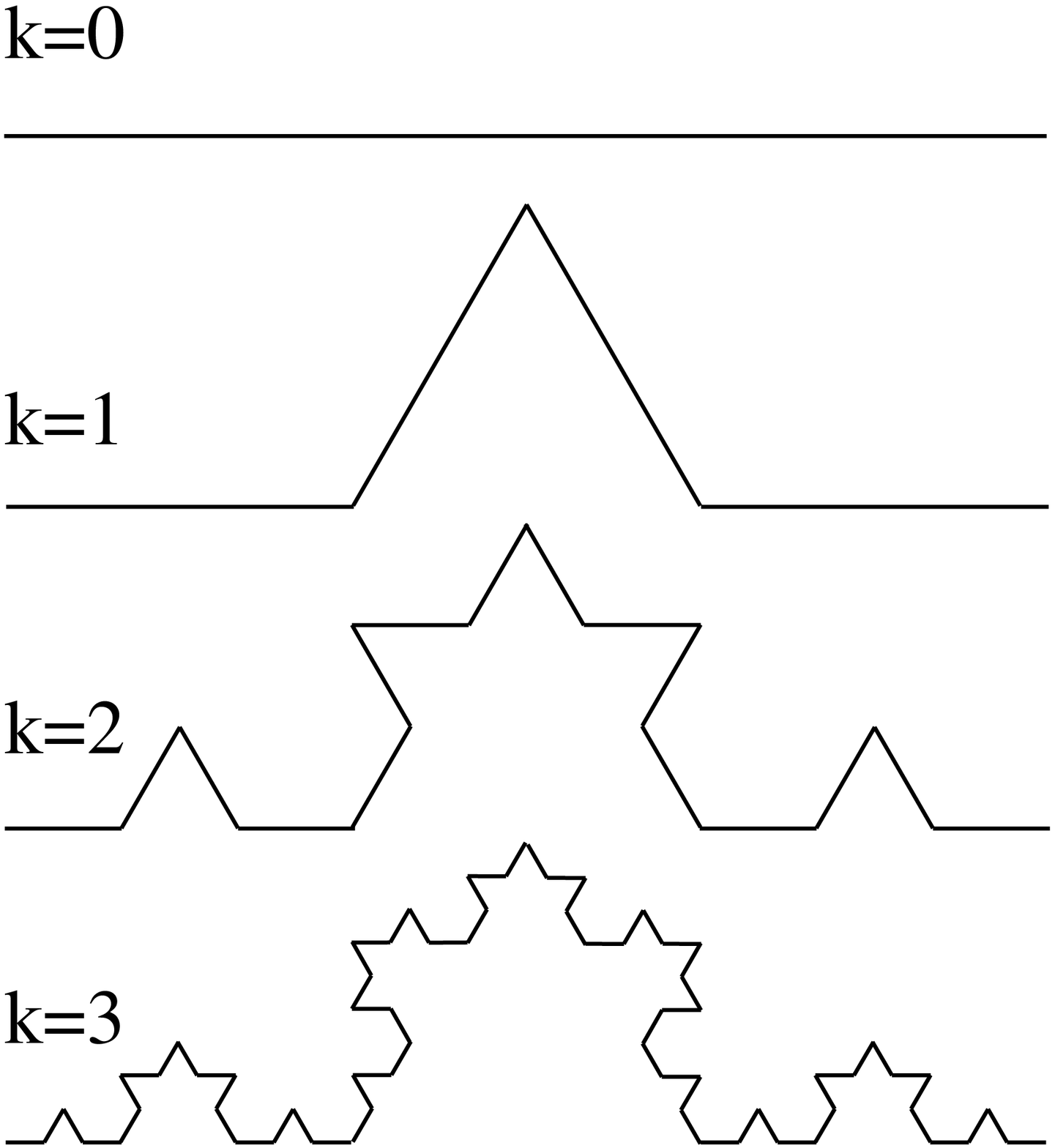, width=8cm}} \vspace*{0.5cm}
  \MycapS{FIG 1. First four iterations in the construction of a van
    Koch curve with $N=4$ and $r=1/3$, $D=\ln 4 / \ln 3 = 1.26186$.}
\end{figure}

\noindent The occupation numbers fit the power-law form predicted by
the MaxEnt with the imposed informational constraint. By comparing this
expression with the general form~\equ{theoretical1}, we can recognize
the
meaning of the Lagrange multiplier $\beta$: It corresponds to the
fractal dimension $D$ of the set.

By substituting $\ell_k / \varepsilon = r^{(k-m)}$ into
eq~(\ref{partition}), the partition function for the
$m$-th iteration reads
\begin{equation}%
Z_m =  \sum_{k=0}^{m} \exp \Bigl( -\beta (k-m)\ln r \Bigr)
\equiv \sum_{k=0}^{m} \exp \Bigl( -\beta k \ln (1/r) \Bigr).
\label{partition1}
\end{equation}%
For the Van Koch curve, the information constraint~\equ{infor} is
expressed as a condition on the average information in a canonical
ensemble given by an
``spectrum'' of information levels
${\cal I}_k = k \ln (1/r)$, an inverse temperature $\beta=D$,  and a
partition function~\equ{partition1}. Its statistical properties are
therefore
formally analoguos to those of the quantum harmonic oscillator,
with a spectrum of energy levels $E_n = \hbar \omega (n +
\frac{1}{2})$ \cite{landsberg78}. The information levels $\cI_k$ are
equally spaced
and they represent the value of the information in each of the
iterative steps leading to the van Koch curve. At each of these steps
the information is increased by a constant quantity $\Delta \cI =
\cI_{k+1} - \cI_k = \ln (1/r)$. This increment stands for the amount of
information needed to push the interation from one level to the next.
The fact that the increment of information is constat allows a formal
computation of the Lagrange multiplier $\beta$ (the fractal dimension)
as a function of the actual information content $\bar{I}$. Consider the
limit $m\to\infty$ in eq.~\equ{partition1}. We obtain the partition
function for the final and strictly self-similar Van Koch curve, namely 
\begin{equation}%
Z(\beta) = \lim_{m\to\infty} Z_m = \frac{1}{1 - e^{-\beta\ln (1/r)}}.
\label{part1}
\end{equation}%
The average information~\equ{infor} is then equal to
\begin{equation}%
\left< I \right> = \frac{\Delta \cI}{e^{\beta \Delta \cI} - 1},
\end{equation}%
where we have introduced $\Delta \cI = \ln(1/r)$. 
The knowledge of the actual value of the average information content
$\bar{I}$  yields finally to the following closed expression
for the fractal dimension:
\begin{equation}%
D = \beta = \frac{1}{\Delta \cI} \ln \left( 1 + \frac{\Delta
\cI}{\bar{I}} \right).
\label{dim1}
\end{equation}%
That is, we have been able to deduce an exact expression for the
fractal dimension $D$, defined as the Lagrange multiplier $\beta$ in
eq.~\equ{theoretical1}, which involves only the informational
magnitudes of the iterative process leading to the fractal pattern.

\section{Two-level iterative model}

The former model can be extended in order to enclose a
more general sort of iterative processes. Let us consider a two-level
process $\cP^*$, in which each iteration order $k$ is composed of
elements belonging to different {\em classes} (which we label with an
index $i$) in a number $n_k(i)$; that is, there are $n_k(i)$ elements
of class $i$ within order $k$.  Each one of these element are
exclusively characterized by a certain magnitude $\ell_k(i)$.  We
define our structural probabilities as the probability of a given
element being of class $i$ and belonging to order $k$, namely $p_k(i) =
n_k(i) / \sum_{k', i'} n_{k'} (i')$.  Similarly, we define the
probability of an element being of class $i$ {\em given that} it
belongs to order $k$  by $p(i/k)=n_k(i) / \sum_{i'} n_k(i')$, and the
probability on an element belonging to order $k$ {\em irrespective of}
its class by $p(k)=\sum_{i'} n_k(i') / \sum_{k',i'} n_{k'}(i')$.  These
distributions trivially fulfil the relation $p_k(i)=p(k) p(i/k)$.
Given our structural probabilities $p_k(i)$, $\cP^*$ has
assigned a total entropy
\begin{equation}%
H(\cP^*)= - \sum_{k,i} p_k(i) \ln p_k(i).
\label{entro2}
\end{equation}%
In order to determine the informational constraint, consider an element
of order $k$ and class $i$, characterized by a value
$\ell_k(i)$. If it is composed of $\ell_k(i)/\ep$
indistinguishable atoms of size $\ep$, then we associate to it an
information content $I_k(i) = \ln \left( \ell_k(i) / \ep \right)$.  An
average over classes $i$ provides the average conditional generating
information of level $k$
\begin{equation}%
\left<I_k\right>= \sum_i p(i/k) I_k(i).
\end{equation}%
A subsequent average over orders yields the global average
information of $\cP^*$

\end{multicols}

\widetext

\Raya\hfill

\begin{equation}%
\left<I\right>= \sum_k p(k) \left<I_k\right> = \sum_k p(k) \sum_i
p(i/k) I_k(i) = \sum_{k,i} p_k(i)  \ln \left( \frac{\ell_k(i)}{\ep}
\right).
\label{lligam2}
\end{equation}%

\hfill\Raya

\begin{multicols}{2}

\noindent Following analogous steps to those in the one-level construction, the
maximization of the entropy~\equ{entro2}, subject to the constraint of
average information~\equ{lligam2} constant, yields the most likely
distribution of occupation numbers
\begin{equation}%
p_k(i) \sim n_k(i) = \mbox{\rm const.} \times \left(
\frac{\ell_k(i)}{\ep} \right)^{-\beta},
\label{theoretical2}
\end{equation}%
where $\beta$ is again a Lagrange multiplier. 

We define a new partition function 
\begin{equation}%
Z(\beta) =  \sum_{k,i} \left( \frac{\ell_k(i)}{\varepsilon}
\right)^{-\beta} = 
\sum_{k,i} \exp \left( - \beta \ln \left(\frac{\ell_k(i)}{\varepsilon}
\right) \right),
\end{equation}%
from which the average information is again given by
eq.~(\protect\ref{infor}). The two-level iterative process is now
analogous to a
canonical ensemble with partition fuction $Z(\beta)$, inverse
``temperature'' $\beta$, and a spectrum of information levels ${\cal
I}_k(i)
= \ln \bigl( \ell_k(i) / \varepsilon  \bigr)$.

\begin{figure}[t]
  \centerline{\epsfig{file=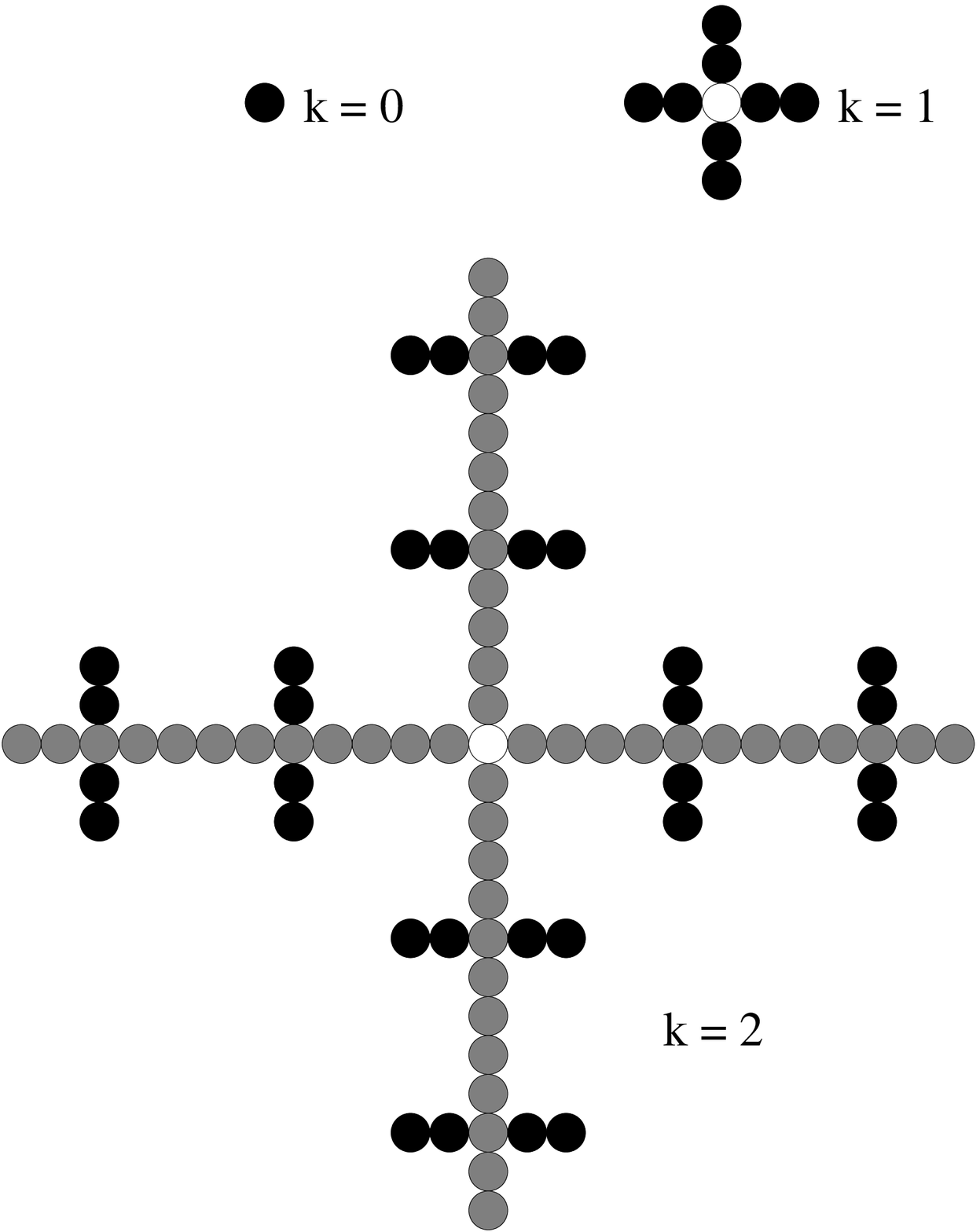, width=8cm}} \vspace*{0.5cm}
  \MycapS{FIG 2. First three iterations in the construction of a
    generalized Vicsek set with $c=2$, $N=9$, and $R=5$, $D=\ln 9 /
    \ln 5 = 1.36521$.}
\end{figure}

As an example of this new model, we propose the
construction of a generalized  Vicsek set
\cite{pastor96,vicsek87}, defined as follows. The
construction of this fractal starts from a seed (particle) of mass
$1$ ($k=0$) and continues in stage $k$ by adding to the previous
($k-1$) structure $4c$ identical copies of it, evenly distributed along
each one of the four main branches of the set. Figure~2 depicts the
generalized Vicsek set corresponding to $c=2$. At each iterative step
the mass (number of particles) of the object increases by a factor
$N=4c+1$, and its linear length by $R=2c+1$, hence the dimension of
the limit fractal set is $D = \ln N / \ln R$
\cite{mandelbrot82}.  Following the same argument as for the original
Vicsek set \cite{pastor96}, every iteration order of the generalized
construction can be decomposed into a set of {\em branches}
characterized by a different mass (number of particles) $s$.  As usual
\cite{pastor96} a branch is defined by the unique continuous path that
starts at the tip of the branch and ends either on the seed or on
another brach of different
mass.  It is easy to check that the branches at each iteration level
can be classified in classes with a different mass $s_k(i)$ and an
occupation number $n_k(i)$ (number of branches of mass $s_k(i)$ in the
$k$-th iteration) given by
\begin{equation}%
n_k(i) = 2 (N-1) N^{i-1}, \quad s_k(i)= \frac{R^{k-i} -1}{2} 
\label{n-de-vicsek}
\end{equation}%
for $i=1,\ldots,k-1$.By removing the $i$ dependence in both
relations~\equ{n-de-vicsek}, we 
obtain
\begin{displaymath}%
n_k(i) = \frac{2(N-1)}{N} N^k \left( 2s_k(i) + 1 \right)^{-D}.
\end{displaymath}%
If we identify the magnitude characterizing branches as $\ell_k(i)
= 2 s_k(i) + 1$, then this expression fits again the prediction of
MaxEnt.

By introducing into the average~\equ{lligam2} the actual
expressions $p_k(i) = N^{i-1} / \sum_{k'} \sum_{i'} N^{i'-1}$ and
$\ell_k(i)=R^{k-i}$ (taking $\ep=1$ the mass of the initial
configuration), and extending the sum over orders to $k=2,\ldots,m$ (to
avoid problems at $k=1$) and over classes to $i=1,\ldots,k-1$, we
obtain after some algebra
\begin{eqnarray}
\left<I\right> & = & \sum_{k=2}^{m} \sum_{i=1}^{k-1}
p_k(i) \ln \left( \frac{\ell_k(i)}{\ep}
\right) = \nonumber \\
& = & \ln R \frac{1}{\sum_{k=2}^{m} \sum_{i=1}^{k-1} N^{i-1}}
\sum_{k=2}^{m} \sum_{i=1}^{k-1} (k-i) N^{i-1} = \nonumber \\
& = & \frac{1}{2} \ln R \frac{1}{\sum_{k=1}^{m-1}  k N^{-k}}
\sum_{k=1}^{m-1}  k (k + 1) N^{-k}. \nonumber
\end{eqnarray}
By defining $Z_m^* = \sum_{k=1}^{m-1}  k N^{-(k+1)}$, we have
\begin{equation}
\left<I\right>  = - \frac{1}{2} \ln R \frac{1}{Z_m^*}
\left( N \frac{d}{d N} \right) Z_m^* =
- \frac{1}{2} \frac{\partial}{\partial \beta} \ln Z_m^*,
\label{canonic2}
\end{equation}
where we have introduced the inverse temperature $\beta=D=\ln N / \ln
R$. The partition function is rewritten
\begin{equation}%
Z_m^* = \sum_{k=1}^{m-1} k N^{-(k+1)} = \sum_{k=2}^{m} (k-1)
\exp \Big(- \beta k \ln R \Big).
\label{partition2}
\end{equation}%
That is, the process $\cP^*$ behaves now, except for an irrelevant
factor $\frac{1}{2}$, as a canonical ensemble given by the partition
function~\equ{partition2}, with ``temperature'' $1 / \beta = 1/ D$ and
a ``spectrum'' of equally spaced information levels $\cI_k = k \ln R$,
but now with a {\em degeneracy} $\Omega_k = k-1$. This degeneracy
corresponds to the effect of the population of classes, in a number
$\Omega_k$, inside each iteration level $k$.

Again we can express $D$ in terms of purely informational magnitudes.
Taking the limit $m\to\infty$ in eq.~\equ{partition2},
\begin{equation}%
Z^*(\beta) = \lim_{m\to\infty} Z_m^* = \frac{1}{(e^{\beta \ln R}
-1)^2},
\end{equation}%
and introducing this expression in eq.~\equ{canonic2}, we have
\begin{equation}%
\left< I \right> = \frac{\ln R}{1 - e^{-\beta \ln R}}.
\end{equation}%
The increment between information levels is now $\Delta \cI = \ln R$,
 so we can write
\begin{equation}%
\left< I \right> = \frac{\Delta \cI}{1 - e^{-\beta \Delta \cI}}.
\end{equation}%
The constraint $\left< I \right> = \bar{I}$ leads finally to the
following
expression for $D$:
\begin{equation}%
D = \beta = \frac{-1}{\Delta \cI}  \ln \left( 1 -
\frac{\Delta \cI}{\bar{I}} \right).
\label{dim2}
\end{equation}%
Eqs.~\equ{dim1} and~\equ{dim2} can be mapped onto the single
expression
\begin{equation}%
D = \frac{1}{\xi \Delta \cI}  \ln \left( 1 +
\frac{\xi \Delta \cI}{\bar{I}} \right),
\label{dim-def}
\end{equation}%
where $\xi=+1$ for the one-level (non-degenerate) model and $\xi=-1$
for the two-level (degenerate) model. The analytic expression of $D$
depends of course on the concrete details of the model considered.
However, both
models render the same result in the limit $|\frac{\xi \Delta
\cI}{\bar{I}}| \ll 1$, namely
\begin{equation}%
D \simeq \frac{1}{\bar{I}}.
\end{equation}%
In this limit we indeed recover an informational version of the well-known
equipartition theorem \cite{landsberg78},  relating the average
information $\bar{I}$ with ``temperature'' $1/\beta = 1/D$.

\section{Conclusions}

In this paper we have shown
how the self-similarity of a wide class of deterministic (iteratively
constructed) fractal sets can be inferred via the Maximum Entropy
principle. The fit is achieved by imposing a constraint concerning the
amount of average information required to specify the structure of the
set (generating information).  In this view, self-similarity arises
from a variational principle, and fractal systems can therefore be
associated, through equations~\equ{infor} and~\equ{canonic2}, with a
canonical ensemble from statistical mechanics. In this ensemble the
role of the temperature $\beta$ is played by the fractal
dimension.  This statistical analogy allows for a completely new
interpretation of fractal sets and fractal dimension in which
deterministic fractals are defined as
systems satisfying a constraint of constant average generating
information, where this latter refers to the amount of knowledge needed
to recover the complete system. A variational principle (MaxEnt)
yields the scaling behavior
characteristic of fractal sets. The fractal dimension is defined as the
Langrange multiplier introduced in the maximization procedure, and it
can be analytically computed if the actual value $\bar{I}$ of the
information is known. A particularly attractive point in our proposal
is that we can define
fractal objects and fractal dimension from first principles, with no
explicit reference to the space-filling properties of the sets. This
could provide a new and promising framework for the study of fractals
with no geometrical counterpart.

Our conclusions can be formally extended to
enclose the more usual random fractals, which in this view are
systems characterized with a constant average generating information.
The stability of natural fractal structures would then be  ensured by
the variational principle from which they came. Moreover, our proposal
is a non-trivial preliminary 
hint of why self-similarity is so frequently seen in Nature; that is to
say, as a trend resulting from an economical way of reaching stability.

%
%

\end{multicols}

\end{document}